\documentclass[11pt,twoside]{article}
\usepackage{asp2014}

\aspSuppressVolSlug
\resetcounters

\begin{document}

\title{The Next Generation of the Montage Image Mosaic Toolkit.}
\author{G. Bruce Berriman,$^1$ J. C. Good,$^2$ B. Rusholme,$^2$ and T. Robitaille$^3$}
\affil{$^1$California Institute of Technology, Pasadena, CA 91125, USA; \email{gbb@ipac.caltech.edu}}
\affil{$^2$California Institute of Technology, Pasadena, CA 91125, USA}
\affil{$^3$ Max Planck Institute for Astronomy, D-69117 Heidelberg, Germany} 

\paperauthor{G. Bruce Berriman}{gbb@ipac.caltech.edu}{orcid.org/0000-0001-8388-534X}{California Institute of Technology}{IPAC}{Pasadena}{CA}{91125}{USA}
\paperauthor{J. C. Good}{jcg@ipac.caltech.edu}{  }{California Institute of Technology}{IPAC}{Pasadena}{CA}{91125}{USA}
\paperauthor{B. Rusholme}{rusholme@ipac.caltech.edu}{  }{California Institute of Technology}{IPAC}{Pasadena}{CA}{91125}{USA}
\paperauthor{T. Robitaille}{robitaille@mpia.de }{  }{Max Planck Institute for Astronomy}{ }{Heidelberg}{ }{D-69117}{Germany}

\begin{abstract}
The scientific computing landscape has evolved dramatically in the past few years, with new schemes for organizing and storing data that reflect the growth in size and complexity of astronomical data sets. In response to this changing landscape, we are, over the next two years, deploying the next generation of the Montage toolkit  ([ascl:1010.036]). The first release (October 2015) supports multi-dimensional data sets ("data cubes"), and insertion of XMP/AVM tags that allows images to "drop-in" to the WWT.  The same release offers a beta-version of  web-based interactive visualization of images; this includes wrappers for visualization in Python. Subsequent releases will support  HEALPix (now standard in cosmic background experiments); incorporation of Montage into package managers (which enable automated management of software builds), and support for a library that will enable Montage to be called directly from Python.This next generation toolkit will inherit the architectural benefits of the current engine - component based tools, ANSI-C portability across Unix platforms and scalability for distributed processing. With the expanded functionality under development, Montage can be viewed not simply as a mosaic engine, but as a scalable, portable toolkit for managing, organizing and processing images. 

The architectural benefits of Montage provide considerable flexibility to the end user, and we will describe how the community is taking advantage of it to integrate its components  into pipelines and workflow environments. Examples include: underpinning a pipeline to create three color SDSS mosaics for galaxies in the RC3 catalogs;  integration into the AAO/UKST SuperCOSMOS H-alpha Survey flux calibration pipeline (Frew et al. 2014, MNRAS, 440, 1080); integration into the processing environment of the Sydney-AAO Multi-object Integral (SAMI) field spectrograph pilot survey; and integration into the processing environment for the Palomar Transient Factory. In addition, it is an exemplar tool for the development of cyberinfrastructure systems that will enable non-experts to  run workflows at scale. One example is building AstroTaverna workflows with Virtual Observatory services.  Another is the production, in collaboration with ISI/USC and Amazon Web Services, of a 16-wavelength Atlas of the Galactic Plane with Open Source tools such as the Pegasus Workflow management system which, when complete, is aimed at deploying a set of  tools for scientists to process and manage data on distributed platforms.

\end{abstract}

\section{Introduction}

In the past ten years, the astronomical data and computational landscapes have evolved rapidly.  Multi-dimensional data sets (data cubes) contain measurements from new complex instruments such as Integral Field Spectrographs. New sky-partitioning schemes are in wide use.  HEALPix organizes images from cosmic background experiments and TOAST presents data in the immersive environment of the WorldWide Telescope ({\footnotesize\url{http://www.worldwidetelescope.org/}}). Python has become the language of choice for many astronomers, in large part due to the success of community packages such as Astropy ({\footnotesize\url{http://www.astropy.org/}}) and its adoption by major projects such as LSST and JWST.  The growth in the volume of data sets has led to a demand for high-performance processing capabilities that operate on remote data and for tools that allow astronomers to take advantage of them. In response to this changing landscape, the Montage Image Mosaic Toolkit has since September 2014 been undergoing a major two-year upgrade.

\section{Deliverables and Release Schedule}
The deliverables and release schedule for the upgrade are summarized in Table 1.  Version 4.0 was released on October 1, 2015 with a BSD 3-clause license. Work has already begun on the deliverables for Version 5.0. In particular, an initial set of tools has been developed for processing and managing data at scale, as part of a project to deliver 16-wavelength Atlas of the Galactic Plane; see   {\footnotesize\url{http://www.noao.edu/meetings/bigdata/files/Berriman.pdf)}}. This paper will discuss two items released in Version 4.0: processing of data cubes and command-line visualization of large FITS files.

\begin{table}[!ht]
\caption{Release Schedule for The Next Generation of Montage}
\smallskip
\begin{center}
{\small
\begin{tabular}{lll}  
\tableline
\noalign{\smallskip}
Version & Release Date & Contents\\
\noalign{\smallskip}
\tableline
\noalign{\smallskip}
Version 4.0 & October 2015 & Data cubes \\
    &   & Command-line visualizer\\
   &    & Python wrapper for visualizer (beta)\\
   &    & Indexing and fast spatial searches for large image data sets \\
\noalign{\smallskip}
\tableline 
\noalign{\smallskip}
Version 5.0 &  October 2016 & HEALPix, TOAST\\ 
  &   &  C library \\
  &   &  Indexing of large data sets \\
  &    & Tools for processing at scale \\
 
\noalign{\smallskip}
 \tableline
\end{tabular}
}
\end{center}
\end{table}
\noindent 

The Version 4.0 distribution may be downloaded from the Montage web page or from GitHub at  {\footnotesize\url{https://github.com/Caltech-IPAC/Montage}}. The release is backwards compatible with earlier releases, retains all the architectural benefits of the toolkit design, and is written in ANSI-C for portability. Formal testing was performed on a RedHat Enterprise Linux Server and on Mac OS X 10.9 platforms. The primary test data set was the Galactic Arecibo L-band Feed Array HI (GALFA-HI) survey \citep{2011ApJS..194...20P}. GALFA is a high-resolution (4'), large-area (13,000 deg$^{2}$), high spectral resolution (0.18 km s$^{-1}$), and wide band (-700 km s$^{-1}$ < v$_{LSR}$ < +700 km s$^{-1}$) survey of the Galactic interstellar medium in the 21-cm  transition of neutral hydrogen, conducted at Arecibo Observatory.

\section{Aggregating Data Cubes Into Mosaics} 
\subsection{Modules for Processing Cubes}
Table 2 lists five new core processing modules dedicated to data cubes. The first four are extensions of the corresponding modules for processing two dimensional images.

\begin{table}[!ht]
\caption{Dedicated Cube Processing Modules in Montage Version 4.0}
\smallskip
\begin{center}
{\small
\begin{tabular}{ll}  
\tableline
\noalign{\smallskip}
Module & Function \\
\noalign{\smallskip}
\tableline
\noalign{\smallskip}
mProjectCube & Reproject data cube (4 dimensions) \\
  mAddCube  & Co-adds reprojected cubes  \\
   mShrinkCube &    Averages data in spatial or physical dimensions \\ 
  mSubCube &  Creates cutouts of cubes  \\
   mTranspose &   Transpose axes of data cubes   \\
\noalign{\smallskip}
\tableline\
\end{tabular}
}
\end{center}
\end{table}
\noindent 

\subsection{How To Create Mosaics of Data Cubes}

The steps in creating a mosaic are analogous to those in creating a two-dimensional mosaic, except that background rectification is not performed. The steps described below
are adapted from a tutorial to create a mosaic of five GALFA images. The complete tutorial takes 9 minutes on a quiet 2.3-GHZ Mac OS X 10.9 machine (see {\footnotesize\url{http://montage.ipac.caltech.edu/docs/cubemosaicstutorial.html}}). The first step, averaging the number of frequency planes from 2048 to 204, is intended simply to reduce processing time:

\begin{enumerate}

\item Average 10 velocity planes to reduce processing time:

\texttt{\small{mShrinkCube -m 10  GALFAinput.fits shrunk/GALFA-shrunk.fits 1}}

\item Create an image list and metadata table of the shrunken images:

\texttt{\small{mImgtbl shrunk images-narrow.tbl}}

\item Create a header template  for the mosaic from the metadata in the image list:

\texttt{\small{mMakeHdr images-narrow.tbl narrow.hdr}}

\item Reproject the  image:

\texttt{\small{mProjectCube GALFAinput.fits proj/GALFAproj.fits template-narrow}}

\item  Create an image metadata file for the reprojected images:

\texttt{\small{mImgtbl proj/proj-narrow.tbl}}

\item Co-add the reprojected images: 

\texttt{\small{mAddCube -p proj/ proj-narrow.tbl narrow.hdr mosaic.fits}}

\end{enumerate}

\subsection{Performance of Data Cube Processing on Amazon Web Services}

To evaluate the new modules at scale and to further evaluate tools for end users, the GALFA data set was mosaicked at full resolution on Amazon Web Services, with an Education Credits Grant awarded through the  AWS SKA AstroCompute Program. A preliminary mosaic was produced in 5.5 hours, with the configuration summarized in Table 3. Further processing passes will optimize the process, and at the end of the project, we will release the product and the scripts used to generate it.

\begin{table}[!ht]
\caption{Processing Summary for Computing GALFA Cubes on the Amazon Web Services Platform}
\smallskip
\begin{center}
{\small
\begin{tabular}{ll}  
\tableline
\noalign{\smallskip}
Module & Function \\
\noalign{\smallskip}
\tableline
\noalign{\smallskip}
Walltime: &  5.5 hours \\
   Resources &  CfnCluster: 5 of m4.2xlarge (1 head node (on-demand); \\
  &  4 workers (spot), 8 vcpu, 32 GB ram each) \\
 Input   &    184 fields, 8.53$^{\circ}$ on a side (114 GB) \\ 
 Output   &  10 fields 35 x 39.3$^{\circ}$, offset 30$^{\circ}$ RA ( 941 files, 857 GB).  \\
   Price
&  us-west-2a:Hourly rate \$0.504  on demand, \$0.0821 spot     \\
 &  Processing cost: \$5  \\
\noalign{\smallskip}
\tableline\
\end{tabular}
}
\end{center}
\end{table}
\noindent

\section{Visualization tools}
A new module, mViewer, is a command line tool that creates PNG or JPEG versions of FITS images, including individual planes in data cubes, with user control over stretch,
coordinate grids, astronomical source (catalog) display with scaled symbols, image metadata outlines, markers (individual symbols),  and labels. It uses
sticky directives that remain in force until new values are specified. 
Here is an an example call to overlay 2MASS sources on a mosaic of M51, derived from SDSS g-band data:

\smallskip
\noindent
\texttt { \small{mViewer  -color ffff00  -symbol 1.0 circle  -scalecol j16.0 mag}} \\
\texttt { \small{-catalog fp-2mass.tbl  -gray SDSSg.fits 0s max  gaussian-log}} \\
\texttt { \small{- out catalog.png}}
\smallskip

\acknowledgements Montage is funded by the National Science Foundation under Grant Number ACI-1440620. Processing on Amazon Web Services is underwritten by an
 Education Credits Grant awarded through the  AWS SKA AstroCompute Program. 

\bibliography{adassXXVreferences} 


\end{document}